\documentstyle[12pt]{article}
\begin{document}
\baselineskip=24pt plus 2pt

\vspace{20mm}

\begin{center}
{\large \bf  A Study of Anyon Statistics by
Breit Hamiltonian Formalism}\\
\vspace{15mm}

G.L. Huang
\vspace{3mm}
\\
Department of Physics,\\
National Cheng Kung University, \\
Tainan, 701, Taiwan\\
Republic of China. \\

\vspace{5mm}

and

\vspace{5mm}
C.R. Lee
\vspace{3mm}
\\
Institute of Physics,\\
National Chung Cheng University, \\
Chia-Yi, 62117, Taiwan,\\
Republic of China. \\

\vspace{20mm}
\hfill {PACS:05.30.-d, 12.20.Ds, 11.15.Kc}
\end{center}
\newpage

\begin{center}
{\bf ABSTRACT}
\end{center}

We study the anyon statistics of a $2 + 1$ dimensional
Maxwell-Chern-Simons (MCS) gauge theory
by using a systemmetic metheod,
the Breit Hamiltonian formalism.



\newpage
\section{Introduction}

In fundamental physics, it is important to classify
elementary particles by their quantum characteristics,
espacially the spin of particles.
Phenomenally and theoretically, in the space-time dimensions
equal or greater than $3 + 1$ dimension, there are just
fermions and bosons with half-integer and integer
spin respectively.
In two spatial dimensions, since the universal covering
group, $SO(2)$, is noncompact and the angular momentum
algebra is a commutation algebra, the elementary
particles get arbitrary values of spin and should
obey fractional statistics.
These particles are called ${\it anyons}$~\cite{1}\cite{2}.

In the usual prescriptions of $2 + 1$ dimensional
quantun field theory, one may impose a magnetic fluxtube
upon every charged
particle (charge-fluxtube composites)~\cite{3}\cite{4},
and then, a phase factor, $e^{i\theta}$,
will appears
in an anyon wavefunction ,while two
identical particles are
made to encircle each other.
The $\theta$ is a constant phase angle,
called the anyon phase angle.
This phase could be treated in terms of wilson
loops, and it is closely related to the Aharonov-Bohm(AB)
effect~\cite{5}\cite{6}.
Viewed in this way, the gauge theory may be
approximatelly described by the non-dynamical
Chern-Simons action rather than the Maxwell's one~\cite{3}.
But, in the true charge-fluxtube composites,
the Coulomb interaction will cause
complicated situations.
Therefore, in this paper, we would like to treat the
statistical phase as a AB effect by
imposing a magnetic momentum, ${\bf m}$, on every
charged particle, but avoiding the assumption of
Charge-fluxtube composites~\cite{7}\cite{8}.

The crucial mission of the magnetic momentum
is to redescribe the statistical interaction through an
introduction of a ${\bf m}\cdot{\bf B}$ type interaction.
The magnetic field ${\bf B}$ is aroused
by the other moving charged particles.
One may assume that the field is a extension
version of $3 + 1$ dimensional Biot-Savart's law,
and it will be the following form, in
Heaviside-Lorentz units,
$$
       {\bf B}=-\frac{q}{2\pi c^2}
       \frac{(\dot{\vec \varrho_2} - \dot{\vec \varrho_1})\times
                        ({\vec \varrho_2} - {\vec \varrho_1})}
                       {|{\vec \varrho_2} - {\vec \varrho_1}|^2}
$$
\begin{equation}
                 =-\frac{q}{2\pi c^2}\frac{d}{dt}(\phi_{12})
                 =-\frac{q}{2\pi c^2}\frac{d}{dt}(\phi_{21})~~,
\end{equation}
where $\vec \varrho_1$ and $\vec \varrho_2$
are the position vectors
of particle 1 and 2 with charge $q$ respectively,
and $\phi_{ab}$ is a relative azimuthal angle
between two particles.  Hence,
the ${\bf m}\cdot{\bf B}$ interaction indeed
become a statistical interaction,
and the statistical potential is
\begin{equation}
L_{\theta}=-2{\bf m}\cdot{\bf B}=\left[\frac{q{\bf m}}{c^2}\right]
         \frac{1}{\pi} \dot\phi~~.
\end{equation}
The factor in the bracket is the anyon phase angle $\theta$.
Consequently, the anyon phase can be expressed as
\begin{equation}
        {\it exp}\left[\frac{i}{\hbar}\int L_{\theta}dt\right]
         ={\it exp}\left[i\frac{\theta}{\pi}\delta\phi\right]~~.
\end{equation}
On our approach,
the ${\bf m\cdot B}$ interaction will be introduced naturally
by identifying this interaction with spin-orbit term
appears in Breit Hamiltonian.
The detail will be discussed later.

We arrange this paper as follow:
In next section, we illustrate what
the Breit Hamiltonian is,
and define some conventions of this paper.
In the third section, the
Breit Hamiltonian are calculated and analyzed in
Coulomb gauge for three theories,
${\it i.e.}$ the Maxwell, Chern-Simons and
MCS gauge theories.  Here,
we choose a minimal coupling scheme
for gauge field and Dirac fermions.
It is also shown how one can naturally
draw a ${\bf m\cdot B}$ interaction into,
once it is identified as a spin-orbit
interaction.
The last section,
an electron-positron system is
explained briefly.
In the appendix,
some useful Fourier transformations are exhibited.



\newpage
\section{Breit Hamiltonian:Non-Relativistic Limit
         of Dirac Theory}

In some situations, physicists prefer to deal a
quantum system in non-relativistic limit.  This
corresponds to a higher order correction
of a potential belongs to Schr\"odinger equation.
For instance, for an $U(1)$ gauge theory with Dirac fermions,
the corrected potetial,
up to the order of $\frac{\hbar^2}{c^2}$,
is just the Breit Hamiltonian~\cite{9}.

The starting point is to solve the Dirac equation
in the non-relativistic limit with
next term in the expansion of the relativistic
expression for the kinetic energy.
It leads to a Schr\"odinger-like equation and
a Schr\"odinger wave function
of a free particle $\varphi_{Sch}$
should satisfy the following equation
$$
      H^0\varphi_{Sch}=(E-mc^2)\varphi_{Sch}~~,
$$
\begin{equation}
      H^0=\frac{\vec p^2}{2m}-\frac{\vec p^4}{8m^3 c^2}~~,
\end{equation}
where $H^0$ is a free fermion Hamiltonian, $E$ is the total
energy and $m$ is fermion mass.
We will denote a spinor amplitude of a plane wave
by $\omega$.
It is subject to normalization condition
$$
  \omega^\ast\omega = 1~~.
$$
Therefore, the ``bispinor" amplitude of
a free particle in a plane then can be
expressed in terms of $\omega$, with correction, by
\begin{equation}
        u(p) = \sqrt{2mc^2} \pmatrix{ (1-\frac{\vec p^2}{8m^2c^2})
                                       \omega \cr
                                       \frac{\vec\alpha \cdot
                                             \vec p}{2mc}
                                       \omega               }~~,
\end{equation}
where $\vec\alpha \equiv \hat i + \sqrt{-1} \hat j$.

We consider the minimal coupling theory of abelian gauge field
and Dirac fermion described by the Lagrangian density
\begin{equation}
      {\cal{L}}={\cal L}_A (A_{\mu},\partial_{\mu}A_{\nu})
    +\overline{\psi}(i {\partial\hskip -7pt /}
-\frac{e}{c} {A\hskip -7pt /}-m)\psi~~,
\end{equation}
in which ${\cal L}_A (A_{\mu},\partial_{\mu}A_{\nu})$
associates with the behaviors
of free photons and $\psi$ is the fermion field.
With the help of Eqs.(5) and (6),
a tree level electron-electron scattering amplitude is
\begin{equation}
        M_{fi} = e^2 (\bar u_1' (p_1') \gamma^{\mu} u_1 (p_1))
                      D_{\mu\nu}(q)
                     (\bar u_2' (p_2') \gamma^{\mu} u_2 (p_2))~~,
\end{equation}
where $p_1,~p'_1,~p_2,~p'_2$ and transfer momentum
$q = p'_1 - p_1 = p_2 - p'_2$ are arranged as Fig.1,
$\vec p=1/2\cdot(\vec p_1-\vec p_2)$ and $D_{\mu\nu}$ is the
photon propagator.
Note that the $\omega_i$ is the spinor amplitude of
particle $i$.
\input FEYNMAN
\begin{center}
\begin{picture}(5000,5000)
\drawline\photon[\S\FLIPPEDCURLY](0,0)[4]
\put(500,-2000){$q$}
\drawarrow[\NE\ATBASE](0,-2200)
\drawline\fermion[\NW\REG](0,0)[5000]
\put(-4500,\pbacky){$p'_1$}
\drawarrow[\NW\ATTIP](\pmidx,\pmidy)
\drawline\fermion[\NE\REG](0,0)[5000]
\put(3900,\pbacky){$p_1$}
\drawarrow[\SW\ATTIP](\pmidx,\pmidy)
\drawline\fermion[\SW\REG](0,-4500)[5000]
\put(-4500,\pbacky){$p'_2$}
\drawarrow[\SW\ATTIP](\pmidx,\pmidy)
\drawline\fermion[\SE\REG](0,-4500)[5000]
\put(3900,\pbacky){$p_2$}
\drawarrow[\NW\ATTIP](\pmidx,\pmidy)
\end{picture}
\end{center}
\vskip 3cm
\input epsf
\begin{center}
{\parbox{10cm}{\small Fig 1.
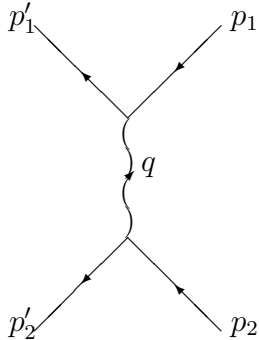
The tree diagram of $e^- e^-$ scattering.}}
\end{center}
\vskip 1cm
In non-relativistic limit, we define the amplitude
\begin{equation}
               M_{fi}
               \equiv -4m^2 c^4 ({\omega_1'}^\ast {\omega_2'}^\ast
                           U(\vec p, \vec q)
                           \omega_1 \omega_2 )~~.
\end{equation}
The $U(\vec p,\vec q)$ is an interacting potential.
By a appropriately Fourier transformation,
\begin{equation}
        \int e^{i\vec q\cdot \vec\varrho}U(\vec p,\vec q)
            \frac{d^2 \vec q }{(2\pi)^2}~~,
\end{equation}
where $\vec\varrho = \vec\varrho_1 - \vec\varrho_2$,
the interaction
potential $U(\vec p,\vec q)$ in equation (8)
is precisely turned out to be the Breit
Hamiltonian in coordinate space which describes the interaction
between two indentical fermions non-relativistically.

In $2 + 1$ dimension,
we defined the complete set of gamma matrice as
\begin{equation}
  \gamma^0\equiv\sigma^3~~,~~~\gamma^1\equiv\sigma^3\sigma^1~~,
  ~~~\gamma^2\equiv\sigma^3\sigma^2~~.
\end{equation}
The $\sigma^i$ are usual Pauli matrice, they satisfy
the following algebra
\begin{equation}
        \gamma^{\mu}\gamma^{\nu} = g^{\mu \nu}
                                   -i\epsilon^{\mu\nu\alpha}
                                               \gamma_{\alpha}~~.
\end{equation}
As space-time indices, all Greek letters take values from 0 to 2,
and all latin letters take values from 1 to 2.
Thereby, from Eqs.(5) and (10), we have
some useful formulae which are needed as scattering
amplitude is handled
and are
exhibited as :
$$
    \bar u'_1 (p'_1) \gamma^0 u_1 (p_1) =
                2mc^2 {\omega_1'}^\ast\left[1
                                       - \frac{\vec q^2}{8m^2 c^2}
                                    + \frac{i\vec q\times\vec p_1}
                                           {4m^2 c^2}\right]
                      \omega_1~~,
$$
$$
        \bar u'_1 (p'_1) \vec{\gamma} u_1 (p_1) =
                 c^2(\frac{1}{c}) {\omega_1'}^\ast\left[
                                       (2\vec p_1 + \vec q)
                                     + i((\vec q)_x\hat j
                                         -(\vec q)_y\hat i)\right]
                                  \omega_1~~,
$$
$$
    \bar u'_2 (p'_2) \gamma^0 u_2 (p_2) =
                2mc^2 {\omega_2'}^\ast\left[1
                                    - \frac{\vec q^2}{8m^2 c^2}
                                    - \frac{i\vec q\times\vec p_2}
                                           {4m^2 c^2}\right]
                      \omega_2~~,
$$
\begin{equation}
        \bar u'_2 (p'_2) \vec{\gamma} u_2 (p_2) =
                 c^2(\frac{1}{c}) {\omega_2'}^\ast\left[
                                  (2\vec p_2 - \vec q)
                                   - i((\vec q)_x\hat j
                                   -(\vec q)_y\hat i)\right]
                                  \omega_2~~.
\end{equation}
We have used the conventions of vector, e.g.
$p_{\mu}$ expressed a 3-vector and $(\vec q)_x$
and $(\vec q)_y$ are
the x and y coponents of 2-vector $\vec q$ respectively.
Throughout this paper we use the conventions
$\epsilon^{012}=1$ and Minkowski metric $g_{\mu\nu}
=diag(+1,-1,-1)$. All of the amplitudes
and Breit Hamiltonians
of this paper
are expressed in terms of
centre-of-mass frame.



\newpage
\section{Anyon Phase Angles}

Our main purpose is to analyze the anyon phase
for a $2 + 1$ dimensional system with an abelian topologically
massive Lagrangian density,
\begin{equation}
           {\cal L}_A (A_\mu,\partial_\nu A_\mu)=
     {\cal L}_{MCS} = -\frac{1}{4} F_{\mu\nu}F^{\mu\nu}
              + \frac{\mu}{2} \epsilon^{\alpha
               \beta\sigma}A_{\alpha}\partial_{\beta}A_{\sigma}~~.
\end{equation}
Before doing this, we first consider two simpler cases,
Maxwell and CS gauge theory.

\begin{flushleft}
\begin{minipage}[t]{100mm}
{\bf 3.1~~Maxwell Gauge Theory}
\end{minipage}
\end{flushleft}

In this case, the free photon Lagrangian density is
set to be
\begin{equation}
       {\cal L}_A (A_\mu,\partial_\nu A_\mu)=
       {\cal L}_{M}=-\frac{1}{4}F_{\mu\nu}F^{\mu\nu}~~.
\end{equation}
Now, the subsequent calculations are considerably simplified
if the photon propagator $D_{\mu\nu}$ is chosen in the
Coulomb gauge.
The non-vanishing components of the propagator in $2 + 1$
dimension are
$$
        D_{00} = -\frac{1}{\vec q ^2}~~,
$$
\begin{equation}
        D_{ij} = \frac{1}{\vec q ^2 }
                 (\delta_{ij} - \frac{(\vec q)_i (\vec q)_j}
                                     {\vec q ^2})~~.
\end{equation}
Note that we take the non-relativistic approximation,
$q^2\simeq \vec q ^2$, throughout this paper.
The corresponding scattering amplitude, with the
help of Eq.(12), is

$$
        M_{ij} = -4m^2 c^4 {\omega_1}'^\ast {\omega_2}'^\ast
                 \left[e^2 \frac{1}{\vec q ^2}
                 - \frac{ e^2}{2m^2 c^2}
                 +i\frac{3 e^2}{2m^2 c^2}\frac{\vec q \times\vec p}
                                             {\vec q ^2} \right.
$$
\begin{equation}
                 \left. +\frac{ e^2 \vec p^2}
                         {m^2c^2}\frac{1}{\vec q^2}
                 -\frac{e^2}{m^2 c^2} \frac{(\vec q \cdot
                                       \vec p)^2}{\vec q ^4}\right]
                  \omega_1 \omega_2~~.
\end{equation}
This formula have been transformed into the type of
centre-of-mass frame.
It is clearly that
particle interaction operator in momentum space version is just
the function of $\vec p$ and $\vec q$ in the bracket of Eq.(16).
Under Fourier transformations, which are displayed in appendix,
one could obtain
the following final expression for Breit Hamiltonian
$$
        U_{M}(\vec p,\vec\varrho) =  -\frac{e^2}{2\pi} \ln(\varrho)
                              -\frac{3e^2 \hbar}{4\pi m^2 c^2}
                                \frac{\vec\varrho\times\vec p}
                                     {\varrho ^2}
                              -\frac{e^2 \vec p ^2}{4\pi m^2 c^2}
                                \ln (\varrho)
$$
\begin{equation}
                              +\frac{e^2}{4\pi m^2 c^2}
                               \frac{1}{\varrho ^2}
                         \vec\varrho\cdot(\vec\varrho\cdot\vec p)
                               \vec p
                              -\frac{e^2\hbar ^2}{2m^2 c^2}
                                \delta^2(\vec\varrho)~~,
\end{equation}
where $\varrho=|\vec \varrho|$ and
$\hbar$ have been put into appropriately.

The term proportional to $\vec\varrho\times\vec p$ in $U_M$
represents spin-orbit interaction.
Since $\vec \varrho\times\vec
p=-\frac{1}{2}m\varrho^2\dot\phi$,
the spin-orbit term,
$
               -\frac{3e^2 \hbar}{4\pi m^2 c^2}
              \frac{\vec\varrho\times\vec p}
{\varrho ^2}~~$,
is identified to ${\bf m\cdot B}$ interaction.
One could pick up an anyon phase angle from the terms like this.
{}From Eq.(17), the anyon phase angle is
\begin{equation}
      \theta_{M}=-\frac{3e^2}{8mc^2}~~.
\end{equation}
If the electron magnetic momentum, ${\bf m}=-\frac{q\hbar}{2mc}$,
is put into Eq.(2), the statistical phase angle would be
\begin{equation}
       \theta=-\frac{q^2}{2mc^2}~~.
\end{equation}
It disagrees with $\theta_{M}$ as $q^2=e^2$.
This is because that the purely kinematical effect,
the Thomas precession effect, $-\frac{q^2}{8mc^2}$,
is absent in the non-relativistic limit.
In fact, these results had been calculated
and analyzed by Hansson, Sporre and Leinass~\cite{7}.
We showed this again just for a purpose
that we could compare these outcomes with the results
of ${\cal L}_{MCS}$.


\begin{flushleft}
\begin{minipage}[t]{100mm}
{\bf 3.2~~Chern-Simons Gauge Theory}
\end{minipage}
\end{flushleft}

Recently a lot of attention has been given to CS
gauge theory in two spatial
dimensions~\cite{10}\cite{11}\cite{12}\cite{13}.
Because it contributed remarkable features
in different topics of physics, e.g.
superconductivity, quantum hall effect,
etc.
That is why we like to study the Breit
Hamiltonian in CS theory.

The CS Lagrangian density is defined as
\begin{equation}
     {\cal L}_A (A_{\mu},\partial_{\nu} A_{\mu})={\cal L}_{CS}=
               \frac{\mu}{2} \epsilon^{\alpha
               \beta\sigma}A_{\alpha}\partial_{\beta}A_{\sigma}~~.
\end{equation}
Two non-zero components of propagator
in the Coulomb gauge are
$$
        D_{0i} = \frac{i}{\mu} \epsilon_{0ij}(\vec q)^j
                  \frac{1}{\vec q ^2}~~,
$$
\begin{equation}
        D_{i0} = -\frac{i}{\mu} \epsilon_{0ij}(\vec q)^j
                  \frac{1}{\vec q ^2}~~.
\end{equation}
Following the same procedures of section 3.1,
the interacting potential is less complicated
than the previous case. It is
\begin{equation}
        U_{CS}(\vec p,\vec q) =- i\frac{2 e^2 \hbar}{\mu mc}
                            \frac{\vec q\times\vec p}{\vec q^2}
                           +\frac{e^2 \hbar^2}{\mu mc}~~.
\end{equation}
By a Fourier transformation, the Breit Hamiltonian becomes
\begin{equation}
        U_{CS}(\vec p,\vec\varrho) = \frac{e^2 \hbar}{\pi\mu mc}
                              \frac{1}{\varrho ^2}
                              \vec\varrho\times\vec p
                             +\frac{e^2 \hbar^2}{\mu mc}
                              \delta ^2(\vec\varrho)~~.
\end{equation}
In this case, coulomb-like potential is absent.
The anyon phase angle is
again given by the $\vec\varrho\times\vec p$
term, and reads
\begin{equation}
        \theta_{CS} = \frac{e^2}{2\mu c}~~.
\end{equation}
This result is the same as the outcome
derived by Haugset and Ravndal from a different
starting point~\cite{14}\cite{15}\cite{16}.
They got a effective partition
function of CS theory by integrating out the
degrees of freedom of the gauge field.
In non-relativistic limit, the effective action equivalent
to a statistical term,
\begin{equation}
      {\bf\it exp}\left[i\frac{\theta}{\pi}\delta\phi\right]~~,
\end{equation}
if the condition,
$\theta = \frac{e^2}{2\mu c}$,
is set up.
This phase angle condition is just the Eq.(24).

If $\mu$ approaches to infinity,
the Dirac fermions would interact with
each other classically.
It means that, these particles
obey a distribution which is close to the Boltzmann type
rather than the usual quantum statistics.
Therefore, the statistical phase angle,
$\theta_{CS}$, is tended to zero as $\mu$ goes to infinity.


\begin{flushleft}
\begin{minipage}[t]{120mm}
{\bf 3.3~~Abelian Topologically Massive Gauge Theory}
\end{minipage}
\end{flushleft}

A $2 + 1$ dimentional topologically massive gauge theory
is governed by a photon Lagrangian density as
$$
      {\cal L}_A (A_\mu,\partial_\nu A_\mu)=
      {\cal L}_{MCS}
      ={\cal L}_{M}+{\cal L}_{CS}
$$
\begin{equation}
                 = -\frac{1}{4} F_{\mu\nu}F^{\mu\nu}
                   + \frac{\mu}{2} \epsilon^{\alpha
                 \beta\sigma}A_{\alpha}\partial_{\beta}A_{\sigma}~~.
\end{equation}
Then the photon propagator, chosen in the
Coulomb gauge, up to order
$\frac{1}{c}$, reads
$$
     D_{00} = \frac{-1}{\vec q^2 + \mu ^2}+O(\frac{1}{c^2})~~,
$$
$$
     D_{0i} =i \mu \frac{1}{\vec q ^2}
            \frac{1}{\vec q^2
            + \mu ^2} \epsilon_{0ij} (\vec q)^j+O(\frac{1}{c^2})~~,
$$
$$
     D_{i0} =-i \mu \frac{1}{\vec q ^2}
            \frac{1}{\vec q^2 + \mu ^2}
               \epsilon_{0ij} (\vec q)^j+O(\frac{1}{c^2})~~,
$$
\begin{equation}
        D_{ij} = \frac{1}{\vec q^2 + \mu ^2}\delta_{ij}
                 - \frac{1}{\vec q^2 }
 \frac{(\vec q)_i (\vec q)_j}{\vec q^2 + \mu ^2}+O(\frac{1}{c})~~.
\end{equation}

After some straightforward calculation,
the interacting potential is
$$
        U_{MCS}(\vec p,\vec q) =  \frac{e^2}{\vec q^2 + \mu ^2}
                             \left[1 + \frac{\mu}{mc} +
                             \frac{\vec p ^2}{m^2c^2}\right]
                            -\frac{e^2}{2 m^2 c^2}
                      \frac{\vec q ^2}{\vec q^2 + \mu ^2}
$$
$$
                     + \frac{i3 e^2}{2 m^2 c^2}
                             \frac{\vec q\times\vec p}
                             {\vec q^2 + \mu ^2}
                           -\frac{i2 e^2 \mu}{mc}
                             \frac{\vec q\times\vec p}
                                  {\vec q^2(\vec q^2 + \mu ^2)}
$$
\begin{equation}
                        -\frac{e^2}{m^2 c^2}
                             \frac{(\vec q\cdot\vec p)^2}
                                  {\vec q^2(\vec q^2 + \mu ^2)}~~,
\end{equation}
where $\hbar$ is not put into yet.
Note that the first order part of $D_{ij}$,
which is denoted by $O(\frac{1}{c})$,
is not contributed to $U_{MCS}$.
This formula manifestly will go back to
the results of sections 3.1 and 3.2
once one take the limit $|\mu|\to 0$ and $|\mu|\to \infty$
respectively.

Using those Fourier transformions in the appendix, the whole
complicated Breit Hamiltonian is
$$
        U_{MCS}(\vec p,\vec\varrho) =\frac{e^2}{2\pi}
                                    K_0 (|\mu|\varrho)
$$
$$
         +  \frac{e^2}{2\pi} K_0(|\mu|\varrho)\left[\frac{\mu}{mc}
                              +\frac{\mu ^2}{2m^2 c^2}
                              +\frac{\vec p ^2}{m^2 c^2}
                             -\frac{1}{m^2 c^2}
                                \frac{1}{\varrho ^2}
                                \vec\varrho\cdot(\vec\varrho\cdot
                                              \vec p)\vec p\right]
$$
$$
             +\frac{e^2}{2\pi} K_1(|\mu|\varrho)\frac{1}{\varrho}
                              \left[-\frac{3\hbar|\mu|}{2m^2 c^2}
                        -\frac{\mu}{|\mu|}\frac{2\hbar}{mc}\right]
                               \vec\varrho\times\vec p
                              +\frac{e^2\hbar}{\pi\mu mc}
                        \frac{1}{\varrho^2}\vec\varrho\times\vec p
$$
$$
                  +\frac{e^2 \hbar ^2}{2\pi |\mu| m^2 c^2}
                               K_1(|\mu|\varrho)\frac{1}{\varrho}
                               \left[\frac{-2}{\varrho ^2}
                               \vec\varrho\cdot(\vec\varrho\cdot
                                             \vec p)\vec p
                               +\vec p^2\right]
$$
\begin{equation}
                  +\frac{e^2 \hbar ^2}{2\pi \mu ^2 m^2 c^2}
                               \frac{1}{\varrho ^2}
                               \left[\frac{2}{\varrho ^2}
                               \vec\varrho\cdot(\vec\varrho\cdot
                                             \vec p)\vec p
                               -\vec p^2\right]
                       +\delta ^2(\vec\varrho)
                       \left[\frac{-e^2\hbar ^2 }{2m^2 c^2}
                         +\frac{e^2\hbar ^2 \vec p^2}
                              {2\mu ^2 m^2 c^2}\right]~~,
\end{equation}
where $K_i(|\mu|\varrho)$ is the modified
Bessel function of the $i-$th rank.

There is something different from previous sections.
Here, we should analyze two physics stages,
the statistics and Breit Hamiltonian,
while the
extreme cases of $\varrho$
are taken.

First, the statistics problem is discussed.
The statistical Lagrangian one form is
\begin{equation}
       L_\theta dt=
         \left[-\frac{3e^2\hbar}{8mc^2}K_1(|\mu|\varrho)|\mu|\varrho
        -\frac{\mu}{|\mu|}\frac{e^2\hbar}{2 c}K_1(\mu\varrho)\varrho
        +\frac{e^2 \hbar}{2\mu c}\right]d\phi~~,
\end{equation}
which is read from $\vec \varrho\times\vec p$ terms.
Since it is not an exact differential form,
the anyon phase is mixed in the path integral of kernel,
\begin{equation}
         {\bf K}(\vec\varrho_f t_f,\vec\varrho_i t_i)
         \sim\int{\cal D}\left[\vec\varrho\right]
        {\it exp} \left[\frac{i}{\hbar}\int_{t_i}^{t_f}
                  \left[L_{\theta}+L_0\right]dt\right]~~,
\end{equation}
where we have separated the whole Lagrangian into statistical one,
$L_{\theta}$, and the rest part,
$L_{0}$.
We denote $i$ and $f$ as initial and final state respectively.
Obviously, the anyon phase can not be factored out as before.

However, it is interesting while a separation
of the two particles is taken to be the limiting cases.
When the separation goes to zero~\cite{17}, the statistical
Lagrangian one form is expressed as
\begin{equation}
        \lim\limits_{\varrho\to 0} L_{\theta}dt
        =-\frac{3e^2\hbar}{8mc^2}d\phi~~.
\end{equation}
Since it becomes an exact differential form,
the anyon  phase could be factored out again.
And, it is just the case of Maxwell gauge theory.
Conversely, the Lagrangian one form would be
\begin{equation}
        \lim\limits_{\varrho\to \infty} L_{\theta}dt
        =\frac{e^2\hbar}{2\mu c}d\phi~~,
\end{equation}
for $\varrho\to\infty$.
This result falls into the case of the CS gauge theory.
These two situations mean that the statistical
effects of Maxwell action
will dominate entire statistical behaviors as the separation
of the two particles run to zero.
And, when the two particles is separated far enough,
the effect of CS action will cover the Maxwell's one.

Next, the asymptotic behaviors of Breit Hamiltonian
are studied.
For $\varrho\to 0$,
the Breit Hamiltonian becomes
$$
  U_{MCS}(\vec p,\vec \varrho) \simeq  -\frac{e^2}{2\pi}\ln (\varrho)
$$
$$
                  -\frac{e^2}{2\pi}\ln (\varrho)\left[
                        \frac{\mu}{mc}
                        +\frac{\mu^2}{2m^2c^2}
                        +\frac{\vec p ^2}{m^2 c^2}
                        -\frac{1}{m^2 c^2}
                         \frac{1}{\varrho ^2}
                         \vec\varrho\cdot(\vec\varrho
                         \cdot\vec p)\vec p\right]
$$
$$
                                     -\frac{3e^2\hbar}{4\pi m^2 c^2}
                                       \frac{1}{\varrho ^2}
                                       \vec\varrho\times\vec p
                                     -\frac{e^2\hbar ^2}{2 m^2 c^2}
                                       \delta ^2 (\vec\varrho)
$$
\begin{equation}
                  +\frac{e^2\hbar ^2 }{2 \mu ^2 m^2 c^2}
                   \delta ^2 (\vec \varrho)\vec p^2~~.
\end{equation}
Except for last term, the second terms with bracket still
have some differences from $U_M$.  For the limit
$\varrho \to \infty$, $U_{MCS}$
turns out to be
\begin{equation}
       U_{MCS}(\vec p , \vec \varrho) \simeq\frac{e^2\hbar}{\pi\mu m c}
                                          \frac{1}{\varrho ^2}
                                          \vec\varrho\times\vec p~~.
\end{equation}
These asymptotic forms of
$U_{MCS}(\vec p,\vec\varrho)$,
Eqs.(34) and (35), are clearly different from
$U_M(\vec p, \vec\varrho)$
and $U_{CS}(\vec p, \vec\varrho)$, but at least,
the main structures
are similar.

Howevre, if we take the extreme values of $|\mu|$ rather than
$\varrho$, the conclusions are also
valid, and the limiting forms of $U_{MCS}$ are
almost the same as Eqs.(34) and (35).
It is easy to check that
the Breit Hamiltonians in momentum space
calculated in sections 3.1 and 3.2
are regained,
as one takes Eq.(28) in the $|\mu|$ limiting situations before the
Fourier transformation is made.
Conversely, if the trasformations are made first,
then the
Breit Hamiltonian in coordinate space
in sections 3.1 and 3.2 will never be recovered.
All these mean, in brief, integration, ${\it i.e.}$
Fourier transformation,
does not commute with the $|\mu|$
limit-taking operation.



\newpage
\section{Conclusions and Discussions}

We have showed that interacting Dirac fermions
in 2 + 1 dimensions behave
as anyons in the non-relativistic limit.
And, we also discussed briefly the asymptotic behaviors
of $U_{MCS}$ with respect to $\varrho$.
{}From the case of ${\cal L}_{MCS}$, one takes a limit
$\varrho \to 0$($\varrho \to \infty$),
will gains almost the same formulae as the ${\cal L}_{M}$
$({\cal L}_{CS})$ gave.
The effects of Maxwell term
dominates while the limit, $\varrho \to 0$,
is taken.
On the other side,
if $\varrho$ approaches to infinity,
then the
Chern-Simons term will dominate.
The reason is when global structures are detected,
topological properties will emerge explicitly.
Since the local observors are always not sensitive to feel
the affections
of topology, the effects of Maxwell's term will overcome the
effects of the topological Chern-Simons term,
while two particles are too closed.
Oppositely, if two particles are separated far enough,
the topological CS's results will suppress the Maxwell's one.

The Hamiltonians (17), (23) and (29) describe the interactions
between two identical fermions.  However, if one of the fermion is
replaced by an antifermion, then,
not only the scattering amplitude but also
the annihilation amplitude, Fig.2, must
take into account~\cite{9}.
\begin{center}
\begin{picture}(9700,9700)
\drawline\photon[\S\FLIPPEDCURLY](-3000,0)[4]   
\put(-2500,-2200){$q$}
\drawarrow[\N\ATBASE](-3000,-2200)
\drawline\fermion[\NW\REG](-3000,0)[5000]
\put(-7700,\pbacky){$p_+$}
\drawarrow[\SE\ATTIP](\pmidx,\pmidy)
\drawline\fermion[\NE\REG](-3000,0)[5000]
\put(800,\pbacky){$p'_+$}
\drawarrow[\NE\ATTIP](\pmidx,\pmidy)
\drawline\fermion[\SW\REG](-3000,-4500)[5000]
\put(-7700,\pbacky){$p'_-$}
\drawarrow[\LDIR\ATTIP](\pmidx,\pmidy)
\drawline\fermion[\SE\REG](-3000,-4500)[5000]
\put(800,\pbacky){$p_-$}
\drawarrow[\NW\ATTIP](\pmidx,\pmidy)
\put(-3400,-10500){(a)}

\drawline\photon[\S\FLIPPEDCURLY](12000,0)[4]   
\put(12500,-2200){$q$}
\drawarrow[\N\ATBASE](12000,-2200)
\drawline\fermion[\NW\REG](12000,0)[5000]
\put(7300,\pbacky){$p'_-$}
\drawarrow[\NW\ATTIP](\pmidx,\pmidy)
\drawline\fermion[\NE\REG](12000,0)[5000]
\put(15900,\pbacky){$p'_+$}
\drawarrow[\NE\ATTIP](\pmidx,\pmidy)
\drawline\fermion[\SW\REG](12000,-4500)[5000]
\put(7300,\pbacky){$p_+$}
\drawarrow[\NE\ATTIP](\pmidx,\pmidy)
\drawline\fermion[\SE\REG](12000,-4500)[5000]
\put(15900,\pbacky){$p_-$}
\drawarrow[\NW\ATTIP](\pmidx,\pmidy)
\put(11700,-10500){(b)}
\end{picture}
\end{center}
\vskip 4cm
\begin{center}
{\parbox{10cm}{\small Fig 2.
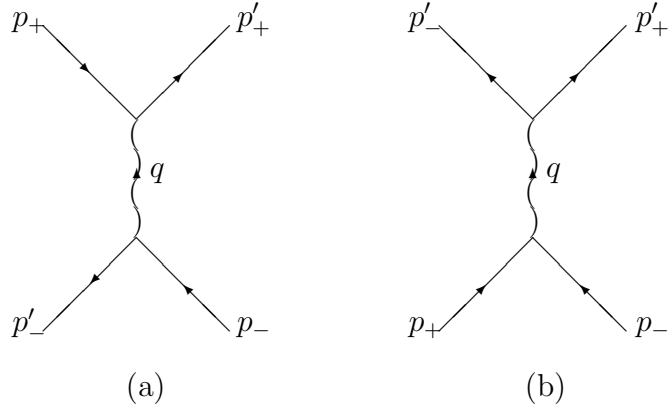
The (a).scattering and (b).annihilation tree diagrams of
an electron-positron system.
The indices ``+" and ``$-$" are defined as
positron and electron respectively.}}
\end{center}
\vskip 1cm

The scattering part is corresponded
to the Hamiltonian (17),
(23) and (29) with opposite sign in centre-of-mass frame, and the
final expression of the annihilation parts
are gotten as :
\begin{equation}
        \cases{ \frac{e^2 \hbar ^2 }{2m^2 c^2} \delta ^2(\vec\varrho)
         \hspace{70mm} {\em for} ~~~~{\cal L}_{M}\cr
                        \cr
                \frac{e^2 \hbar^2}{\mu mc}\delta^2
                                            (\vec\varrho)
         \hspace{72mm} {\em for} ~~~~{\cal L}_{CS}\cr
                        \cr
                \frac{e^2 \hbar^2}{4m^2c^2-\mu ^2}
                     \left[(2-\frac{\mu}{mc})\delta ^2(\vec\varrho)
                -\frac{1}{4m^2c^2} \vec{\nabla}
                      ( \delta ^2 (\vec{\varrho}))\right]
                      ~~~~~~~~~{\em for}~~~{\cal L}_{MCS}      }~~,
\end{equation}
where the result for ${\cal L}_{M}$
enhances the strength of fermions contact term($\delta$-term),
while the outcome for ${\cal L}_{CS}$ will cancel
the term in pure Chern-Simons's one.
Certainly, if $|\mu|\to \infty$,
the last annihilation amplitude will give
the same result as CS gauge theory.
But, as $|\mu|\to 0$, the amplitude does not
approach to the result for ${\cal L}_{M}$.
This ambiguity is again induced by the same reason
for the problem of the asymptotic form of $U_{MCS}$.
It means that, the Fourier transformation does
not commute with the limit-taking operation.

\newpage



\begin{center}
{\large \bf APPENDIX}
\end{center}

In this appendix, some helpful Fourier transformations
in $2 + 1$ dimension would be exhibited as follows~\cite{17}:

\begin{equation}
\int e^{i\vec q\cdot\vec \varrho} \frac{1}
                                  {\vec q ^2}
                       \frac{ d^2\vec q}{(2\pi)^2}
          = -\frac{1}{2\pi}\ln(\varrho)~~,
\end{equation}
\vspace{3mm}
\begin{equation}
\int e^{i\vec q\cdot\vec \varrho} \frac{\vec q}
                                  {\vec q ^2}
                       \frac{ d^2\vec q}{(2\pi)^2}
                 = i\frac{1}{2\pi}
                    \frac{\vec\varrho}{\varrho ^2}~~,
\end{equation}
\vspace{3mm}
\begin{equation}
\int e^{i\vec q\cdot\vec \varrho}\frac{(\vec q\cdot \vec a)
                                           (\vec q\cdot \vec b)}
                                  {\vec q^4}
                                 \frac{d^2 \vec q}{(2\pi)^2}
        = \frac{1}{4\pi}\left[-\ln(\varrho) (\vec a\cdot\vec b)
          -\frac{1}{\varrho ^2}\vec \varrho(\vec\varrho\cdot\vec a)
                                \cdot \vec b \right]~~,
\end{equation}
\vspace{3mm}
\begin{equation}
       \int e^{i\vec q\cdot\vec \varrho}
                   \frac{1}{\vec q^2 + \mu ^2}
                    \frac{d^2 \vec q}{(2\pi)^2}
                   =\frac{1}{2\pi} K_0(|\mu|\varrho)~~,
\end{equation}
\vspace{3mm}
\begin{equation}
  \int e^{i\vec q\cdot\vec \varrho} \frac{\vec q^2}
                                     {\vec q^2 + \mu ^2}
                       \frac{d^2\vec q}{(2\pi)^2}
    = \delta ^2(\vec\varrho)
     - \frac{\mu ^2}{2\pi} K_0(|\mu|\varrho)~~,
\end{equation}
\vspace{3mm}
\begin{equation}
  \int e^{i\vec q\cdot\vec \varrho} \frac{\vec q}
                                    {\vec q ^2 + \mu ^2}
                            \frac{d^2\vec q}{(2\pi)^2}
   = i\frac{|\mu|}{2\pi} K_1(|\mu|\varrho)\hat\varrho ~~,
\end{equation}
\vspace{3mm}
\begin{equation}
  \int e^{i\vec q\cdot\vec \varrho} \frac{1}
                                    {\vec q ^2(\vec q^2 +\mu ^2)}
                                    \frac{d^2\vec q}{(2\pi)^2}
  =  -\frac{1}{2\pi\mu ^2}\left[\ln(\varrho)
     + K_0(|\mu|\varrho)\right]~~,
\end{equation}
\vspace{3mm}
\begin{equation}
  \int e^{i\vec q\cdot\vec \varrho} \frac{\vec q}
                                 {\vec q^2(\vec q^2 + \mu ^2)}
                                 \frac{d^2\vec q}{(2\pi)^2}
 = -i\frac{1}{2\pi|\mu|}
                  K_1(|\mu|\varrho)\hat\varrho
                  +i\frac{1}{2\pi\mu ^2}
                                     \frac{1}{\varrho ^2}
                                     \vec \varrho~~,
\end{equation}
\vspace{3mm}
$$
   \int e^{i\vec q\cdot\vec \varrho} \frac{(\vec q\cdot\vec a)
                                       (\vec q \cdot\vec b)}
                                       {\vec q^2(\vec q^2 +\mu ^2)}
                            \frac{d^2\vec q}{(2\pi)^2}
   = \left[\frac{2}{|\mu|\varrho}K_1(|\mu|\varrho)
                        + K_0(|\mu|\varrho)
                        -\frac{2}{(\mu\varrho)^2}\right]
                \frac{\vec \varrho(\vec \varrho\cdot
                 \vec a)\cdot\vec b}{2\pi\varrho^2}
$$
\begin{equation}
                +\left[-\frac{1}{|\mu|\varrho}K_1(|\mu|\varrho)
                   +\frac{1}{(\mu\varrho)^2}
                   -\frac{\pi}{\mu^2}\delta ^2(\vec\varrho)\right]
         \frac{(\vec a\cdot\vec b)}{2\pi}~~,
\end{equation}
\vspace{3mm}
where $\hat\varrho$ is a unit radial vector in a plane.
The last equation which we would like to call "dipole" term.
Since $\frac{1}{\vec q^2(\vec q^2 + \mu ^2)}$ can
be separated as $\frac{1}{\mu ^2}\left[\frac{1}{\vec q^2}
                   -\frac{1}{\vec q^2 + \mu ^2}\right]$,
and, for example, first part of the separation is looked as
\begin{equation}
    \int e^{i\vec q\cdot\vec \varrho}
          \frac{(\vec a\cdot\vec q)\vec q}{\vec q^2}
          \frac{d^2\vec q}{(2\pi)^2}
        = -\vec \nabla\left[(\vec a)\cdot\vec\nabla
                          (-\frac{1}{2\pi}\ln(\varrho))\right]~~,
\end{equation}
which is just the type of dipole electric field
with respect to a dipole momentum $\vec a$.

\newpage



\vspace{20mm}
\begin{center}
{\large \bf Acknowledgements}
\end{center}
G.L. Huang would like to thank Prof. R.R. Hsu and Prof. S.L. Nyeo
for useful discussions and comments, and Mr. G.J. Cheng for his
kind help.
This research is supported in part by the National Science Council
of the Republic of China under Grant No.~NSC 82-0208-M-194-07.
\newpage



\newpage

\end{document}